\documentclass[prl,twocolumn,amsmath,amsfonts,amssymb,showpacs]{revtex4}
\usepackage{epsfig,psfrag,graphicx,bm}
\begin{document}

\title{Gamma Rays from Kaluza-Klein Dark Matter}

\author{Lars Bergstr\"om}
\email{lbe@physto.se}

\author{Torsten Bringmann}
\email{troms@physto.se}

\author{Martin Eriksson} 
\email{mate@physto.se} 

\author{Michael Gustafsson}
\email{michael@physto.se}

\affiliation{Department of Physics, Stockholm University, AlbaNova
  University Center, SE - 106 91 Stockholm, Sweden}

\date{October 15, 2004}

\begin{abstract}
A TeV gamma-ray signal from the direction of the Galactic center (GC) has been detected by the H.E.S.S. experiment. Here, we investigate whether Kaluza-Klein (KK) dark matter annihilations near the GC can be the explanation. Including the contributions from internal bremsstrahlung as well as subsequent decays of quarks and $\tau$ leptons, we find a very flat gamma-ray spectrum which drops abruptly at the dark matter particle mass. For a KK mass of about 1 TeV, this gives a good fit to the H.E.S.S.~data below 1 TeV. A similar model, with gauge coupling roughly three times as large and a particle mass of about 10 TeV, would give both the correct relic density and a photon spectrum that fits the complete range of data.
\end{abstract}
\maketitle

\newcommand{\B}{B^{(1)}}
\newcommand{\hg}{\hat{g}}
\newcommand{\hgp}{\hat{g}'}
\newcommand{\hv}{\hat{v}}
\newcommand{\tv}{\tilde{v}}
\newcommand{\kko}{{(\!\!\:0\!\!\:)}}
\newcommand{\kkn}{{(\!\!\:n\!\!\:)}}
\newcommand{\kkm}{{(\!\!\:m\!\!\:)}}
\newcommand{\kkk}{{(\!\!\:k\!\!\:)}}
\newcommand{\kkl}{{(\!\!\:l\!\!\:)}}
\newcommand{\pa}{\partial}
\newcommand{\vs}{\!\!\!\!\!}
\newcommand{\bl}{\Big[}
\newcommand{\br}{\Big]}
\newcommand{\nn}{\nonumber}
\newcommand{\de}{\delta}
\newcommand{\al}{\alpha}
\newcommand{\ga}{\gamma}
\newcommand{\ps}{\!\!\not\mbox{\hspace{-0.5mm}}p}
\newcommand{\ks}{\!\!\not\mbox{\hspace{-0.5mm}}k}
\newcommand{\qs}{\!\!\not\mbox{\hspace{-0.5mm}}q}
\section{Introduction}

Currently there is great interest in the very high-energy gamma-ray
signal from the center of the galaxy, where recently the CANGAROO
\cite{cang}, VERITAS \cite{verit} and H.E.S.S. \cite{hess}
collaborations have reported results. The three experiments find
different spectra, something that may depend on different angular
acceptances, energy calibrations, etc. The H.E.S.S. result
superficially looks more detailed and accurate and is therefore the
one we will focus on here, although of course the three groups will
have to find a conclusive reason for the reported discrepancies.
There have already been speculations of a TeV-scale dark matter (DM)
particle giving, through annihilations, a steady source of these
photons \cite{sarkar,hess,horns}. Of course, there are other
possibilities for emission of these gamma rays related to the
physics around the Galactic massive black hole \cite{aharonian}.

The prototype DM particle is the neutralino, the lightest
supersymmetric particle. The phenomenology of this candidate has been
worked out in detail over the last two decades (for a summary, see
\cite{neut}). However, these models give a rather soft gamma-ray
spectrum, so one has to push masses up to the tens of TeV range to
match the results of H.E.S.S., which will be our main concern in this
Letter. We focus instead on an interesting alternative which has
received much attention in recent years; that of universal extra
dimensions (UED) \cite{app,ser,che}. In these models, all standard
model fields propagate in the higher dimensional bulk. 
For the effective four-dimensional theory, this means that all
particles are accompanied by a tower of increasingly more massive 
Kaluza-Klein (KK) states. Momentum conservation in the extra
dimensions, which corresponds to KK mode number conservation, 
is broken by an orbifold compactification which 
projects out unwanted degrees of freedom at the zero mode level. If
the boundary terms introduced at the orbifold fixed points are
identical, then a remnant of KK mode number conservation is left in
the form of KK parity, and the lightest KK particle (LKP) is stable
\cite{che}. This is analogous to conserved R-parity in supersymmetric
models which ensures the stability of the lightest supersymmetric
particle. If the LKP is also neutral and non-baryonic, it is a
potential dark matter candidate.

We consider the simplest, five-dimensional model with one 
compactified UED on an $S^1/Z_2$ orbifold of radius $R$. At tree level,
the $n$th KK mode mass is then given by
\begin{equation}
  m^{(n)} = \sqrt{(n/R)^2 + m_\text{EW}^2},
\end{equation}
where $m_\text{EW}$ is the zero mode mass. However, the identification
of the LKP is nontrivial because radiative corrections to the mass
spectrum of the first KK level are typically larger than the
corresponding electroweak mass shifts given by $m_\text{EW}$. 
A one-loop calculation
\cite{che} shows that the LKP is well approximated by the first KK
mode of the hypercharge gauge boson $\B$. The $\B$ relic density was
determined in \cite{ser}. Depending on the exact form of the mass
spectrum and the resulting coannihilation channels, the limit from the
Wilkinson Microwave Anisotropy Probe (WMAP) \cite{spe} of
$\Omega_\text{CDM} h^2 = 0.12 \pm 0.02$ corresponds to $0.5\text{~TeV}
\lesssim m_{\B} \lesssim 1\text{~TeV}$. Here $\Omega_\text{CDM}$ is
the ratio of DM to critical density and $h$ is the Hubble constant in
units of $100\text{~km} \text{\,s}^{-1} \text{\,Mpc}^{-1}$. Collider
measurements of electroweak observables give a current constraint of
$R^{-1} \gtrsim 0.3\text{~TeV}$ \cite{app,aga}, whereas LHC should
probe compactification radii up to 1.5 TeV \cite{cheb}.

The prospects of KK DM detection have been studied in some detail
\cite{chec,hoo,maj,serb,ber,hoo2}. Secondary gamma rays from
cascading quark decays due to $\B\B$ annihilation in the GC produce
relatively soft spectra \cite{chec,ber}. However, unlike the
supersymmetric case, charged lepton production is not helicity
suppressed, which, e.g., results in a striking peak signal in the
positron spectrum \cite{chec,hoo2} for masses $\lesssim$ 0.5 TeV. This
has led us to also study the decays of $\tau$ leptons and the spectrum
of \emph{primary} continuum gamma rays from the production and
subsequent radiative photon emission of high-energy charged leptons.

\section{Primary and secondary gamma rays from $\B\B$ annihilations}

We first consider primary gamma rays. At tree level, with all other
first level KK modes degenerate in mass, $\B$ pairs annihilate into
quark pairs (35\%), charged lepton pairs (59\%), neutrinos (4\%),
charged (1\%) and neutral (0.5\%) gauge bosons, and Higgs bosons
(0.5\%). These branching ratios agree with those obtained by Servant
and Tait \cite{ser} (however, they do not take into account
electroweak symmetry breaking). In this Letter we only consider
photons radiated from charged leptons $\ell^\pm$. We do not concern
ourselves with the radiative processes which electrons and positrons
encounter during their propagation through the Galaxy, as these give
mostly low-energy photons. The exception may be inverse Compton
scattering on infrared photons and starlight, but generally this is
expected to be only a small correction \cite{aloisio}.

\begin{figure}[htbp]
\psfrag{l+}[][][1]{\mbox{\footnotesize $\ell^+$}}
\psfrag{l-}[][][1]{\mbox{\footnotesize $\ell^-$}}
\psfrag{l1}[r][r][1]{\mbox{\footnotesize $\ell^{\text{{\tiny(1)}}}$}}
\psfrag{b}[][][1]{\mbox{\footnotesize $B^{\text{{\tiny(1)}}}$}}
\psfrag{g}[][][1]{\mbox{\footnotesize $\gamma$}}
\includegraphics[width=0.9\columnwidth]{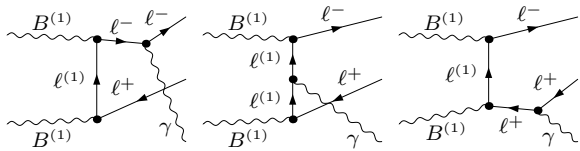}
\caption{Contributions to $\B\B \rightarrow
\ell^+ \ell^- \ga$.}
\label{feyn}
\end{figure}

The tree level Feynman diagrams which contribute to the process $\B\B
\rightarrow \ell^+ \ell^-\ga$ are shown in Fig.~\ref{feyn}. The
computation of the cross section is straightforward and we find that
the differential photon multiplicity is well approximated by
\begin{align}
  \label{sigma}
  \frac{\text{d}N_\ga^\ell}{\text{d}x} & \equiv
  \frac{\text{d}(\sigma_{\ell^+\ell^-\ga}v)/\text{d}x}{\sigma_{\ell^+\ell^-}v}
  \nn \\ & \simeq \frac{\alpha}{\pi} \frac{(x^2 - 2x + 2)}{x} \ln{\left[
  \frac{m^2_{\B}}{m^2_\ell}(1-x) \right]},
\end{align}
where $x\equiv E_\gamma / m_{\B}$. From the electromagnetic coupling
and the phase space difference between two- and three-body final
states, one would expect an answer of the order of $\alpha/\pi$ times
a large logarithm, which is related to a collinear divergence. Our
calculation shows that there is indeed such a leading logarithmic
term, giving large contributions for high photon energies $E_\ga$.
Restricting ourselves to these energies (at lower energy, quark
fragmentation dominates anyway), the radiative correction, although
large, is not more than some five percent of the lowest order cross
section. Therefore, there is no need in this first treatment of the
problem to sum leading logarithms, but this could eventually be done
\cite{bergb}.

We would like to emphasize that our result (\ref{sigma}) is almost
entirely due to the very large mass of the $\B$ and practically
independent of the initial state spin. Expressed in the scaling
variable $x$, it is furthermore quite insensitive to the $\B$ mass and
the mass splitting at the first KK level. This result therefore
applies not only to KK DM, but to any weakly interacting massive
particle (WIMP) candidate with nonsuppressed couplings to charged
leptons.

Let us now consider secondary gamma rays. As already mentioned,
cascading decays of $q\bar{q}$ final states have been studied
previously \cite{chec,ber}. Here, we also include the semihadronic
decays of $\tau$ leptons, which are important in KK DM models since
they have a fairly hard spectrum and a branching ratio of around
20\%. We will use the recent results of Fornengo, Pieri, and Scopel
\cite{for}, who have used the \textsc{Pythia} Monte Carlo code
\cite{sjo} to parametrize $\text{d}N_\ga^{q,\tau}/\text{d}x$ for
quarks and $\tau$ leptons with a center of mass energy of 1 TeV. We
neglect the few percent going into $W$, $Z$, and Higgs final states.

\begin{figure}[t]
\psfrag{y}[][][1.0]{$x^2 \text{d}N_\gamma^\text{eff}/\text{d}x$}
\psfrag{x}[][][0.95]{$x=E_\gamma / m_{\B}$}
\psfrag{1}[][][0.8]{\textsf{1}}
\psfrag{0.1}[][][0.8]{\textsf{0.1}}
\psfrag{0.03}[][][0.8]{\textsf{0.03}}
\psfrag{0.01}[][][0.8]{\textsf{0.01}}
\includegraphics[width=\columnwidth]{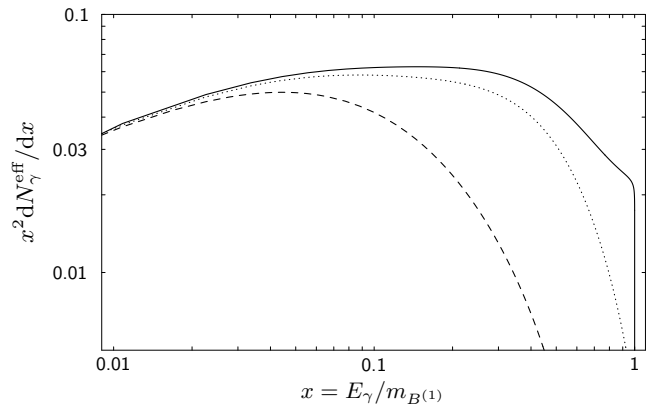}
\caption{The total number of photons per $\B\B$ annihilation (solid
line), multiplied by $x^2=(E_\ga/m_{\B})^2$. Also shown is what quark
fragmentation alone would give (dashed line), and adding to that
the contribution from $\tau$ leptons (dotted line). Here we have assumed
a $\B$ mass of 0.8 TeV and a 5\% mass splitting at the first KK level,
but the result is quite insensitive to these parameters.}
\label{spectrum}
\end{figure}

The total number of photons per $\B\B$ annihilation is given by
$\text{d}N_\ga^\text{eff}/\text{d}x \equiv \sum_i \kappa_i
\text{d}N_\ga^i/\text{d}x$, where the sum is over all processes that
contribute to primary or secondary gamma rays, and $\kappa_i$ are
the corresponding branching ratios. The result is shown in
Fig.~\ref{spectrum}. Previous analyses of the photon flux correspond
to the much softer and sharply falling spectrum from quark
fragmentation alone. Also shown is the more important contribution
from $\tau$ decays, and of course the radiative direct process which
is the main topic of this Letter.

\section{gamma-ray flux from the \mbox{galactic center}}

The details of the Galactic halo profile are to a large extent
unknown. High resolution N-body simulations favor cuspy halos, with
radial density distributions ranging from $r^{-1}$ \cite{nfw} to
$r^{-1.5}$ \cite{moo} in the inner regions. It has 
further been suggested that adiabatic accretion of DM 
onto the massive black hole
in the center of the Milky Way may produce a dense spike of $r^{-2.4}$
\cite{gon}. This has, however, been contested \cite{pierospike,milos}.
On the other hand, adiabatic contraction from the dense stellar
cluster, which is measured to exist near the GC, is likely to compress
the halo DM substantially \cite{primack,klypin}. This means that there
is support for a rather dense halo profile very near the center -
something that may be tested with the new 30-m-class telescopes
\cite{ghez}. Bearing these uncertainties in mind, we will use the
moderately cuspy ($r^{-1}$) profile by Navarro, Frenk and White (NFW)
\cite{nfw}.

Following \cite{berg}, the differential gamma-ray flux from WIMP
annihilation in the GC can be written as
\begin{multline}
  \label{phi}
  E_\ga^2 \frac{\text{d}\Phi_\ga(\Delta \Omega)}{\text{d}E_\ga} \simeq
  3.5\cdot10^{-8}\, x^2 \frac{\text{d} N_\ga^\text{eff}}{\text{d} x}
  \bigg( \frac{\sigma_\text{tot}v}{3\cdot10^{-26} \text{~cm}^3
  \text{\,s}^{-1}} \bigg) \\ 
  \times \left( \frac{0.8\text{~TeV}}{m_{\B}} \right) \langle J_{GC}
  \rangle_{\Delta \Omega}\, \Delta \Omega \text{~m}^{-2}
  \text{\,s}^{-1}\text{\,TeV},
\end{multline} 
where $\sigma_\text{tot}v$ is the total $\B$ annihilation rate, and
$\langle J_{GC} \rangle_{\Delta \Omega}$ is a dimensionless
line-of-sight integral averaged over $\Delta \Omega$, the angular
acceptance of the detector. For a NFW profile, one expects $\langle
J_{GC} \rangle_{\Delta \Omega} \Delta \Omega = 0.13\,b$ for $\Delta
\Omega = 10^{-5}$ \cite{ces}, which is appropriate for the
H.E.S.S. telescope. Here, we allow for an explicit boost factor $b$
that parametrizes deviations from a pure NFW profile ($b=1$) and may
be as high as 1000 when taking into account the expected effects of
adiabatic compression \cite{klypin}. For KK DM, $\sigma_\text{tot}v
\simeq 3\cdot10^{-26} (0.8\text{~TeV}/m_{\B})^2 \text{~cm}^3
\text{\,s}^{-1}$ \cite{ser}.

\begin{figure}[t]
\psfrag{x}[][][0.95]{$E_\gamma$ [TeV]}
\psfrag{y}[][][1.0]{$E_\gamma^2~\text{d}\Phi_\gamma/\text{d}E_\gamma$ [$\text{m}^{-2}\text{\,s}^{-1}\text{\,TeV}$]}
\psfrag{0.1}[][][0.8]{\textsf{0.1}}
\psfrag{1}[][][0.8]{\textsf{1}}
\psfrag{10}[][][0.8]{\textsf{10}}
\psfrag{mmm9}[][][0.8]{$\mathsf{10^{-9}}$}
\psfrag{mmm8}[][][0.8]{$\mathsf{10^{-8}}$}
\psfrag{mmm7}[][][0.8]{$\mathsf{10^{-7}}$}
\includegraphics[width=\columnwidth]{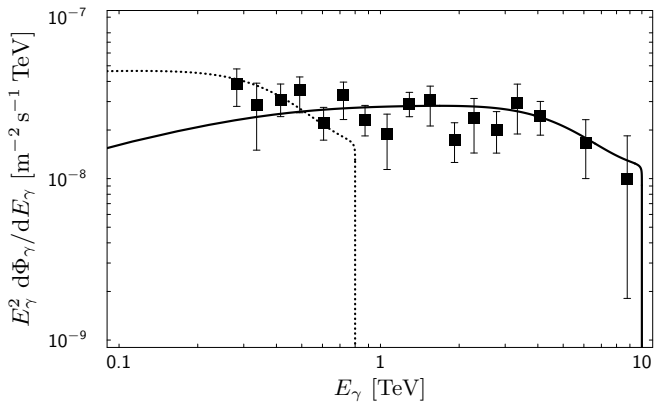}
\caption{The H.E.S.S. data \cite{hess} compared to the gamma-ray
flux from a region of $10^{-5}$ sr encompassing the GC, for a $\B$
mass of 0.8 TeV, a 5\% mass splitting at the first KK level, and a
boost factor $b$ around 200 (dashed line). The solid line corresponds
to a hypothetical 10 TeV WIMP with similar couplings, a total
annihilation rate given by the WMAP relic density bound, and a boost
factor around 1000.}
\label{hess}
\end{figure}  

In Fig.~\ref{hess}, we plot the total differential gamma-ray flux
for a $\B$ mass of 0.8 TeV together with the recent
H.E.S.S. observations with its surprising, relatively hard spectrum
\cite{hess}. As can be seen, an excellent fit to the data is obtained
for energies up to $m_{\B}$, requiring only a moderate value for the
boost factor $b$ of roughly 100. A more detailed analysis would of
course take the energy resolution of the detector into account, giving
a less rapid cutoff at the $\B$ mass.

In order to explain the whole observed spectrum with KK DM, one would
thus need a mass of \mbox{$m_{\B}\gtrsim 7$ TeV}. Such a high mass,
however, is outside the range selected by the relic density constraint
and thus disfavored. It has to be kept in mind, though, that models
with extra dimensions may have quite a different behavior near
freeze-out than ordinary models, which could alter the conclusions
about the preferred mass range. Abnormally large running of the $U(1)$
coupling might also allow for larger masses \cite{die}. 

As remarked before, we expect a similar spectrum for any WIMP dark
matter candidate with nonsuppressed couplings to charged
leptons. Such a model would give a nice fit to the full H.E.S.S. data,
if the canonical WIMP relation given by the WMAP relic density bound,
$(\sigma v)_\text{WIMP}\sim 3\cdot10^{-26} \,\text{cm}^3
\,\text{s}^{-1}$ \cite{neut}, allows for particle masses above 7 TeV. 
For the KK DM scenario considered here, this would require the $\B$'s
gauge coupling to be increased by a factor of roughly three. As an
illustration, we have included such a hypothetical particle of mass 10
TeV in Fig.~\ref{hess} with a boost factor of roughly 1000. Models
which only get a contribution from quark fragmentation, on the other
hand, need a much larger boost factor to fit the observed range. For
supersymmetry \cite{horns}, for example, one is furthermore pushed to
a neutralino mass larger than about 12 TeV, which starts to involve
severe fine-tuning of parameters. It has to be remembered, though,
that the radiative effects investigated in this Letter have not yet
been systematically studied for supersymmetric models \cite{bergb}.

Of course, the models presented in Fig.~\ref{hess} must not violate the
constraints given by the existing EGRET \cite{EGRET}
measurements in the energy range 30 MeV - 10 GeV. We find, however,
that these upper bounds are easily satisfied.
There are some further tests which should be performed in the future to check
the KK hypothesis of the TeV gamma-ray signal. Of course, a signal
from annihilating DM is not expected to show any detectable time
dependence. Since the mass of the DM particle is an abolute cutoff for
the energy of any secondary particle, a steplike drop should
definitely be searched for. If one gets instruments with better energy
resolution, the line from $\B \B \rightarrow \gamma\gamma$
annihilation \cite{bergc} may form the final proof of the particle DM
hypothesis \cite{berg}. It will be interesting to follow the wealth
of new results expected from H.E.S.S., as well as MAGIC \cite{magic},
CANGAROO \cite{cang} and VERITAS \cite{verit}. The GLAST satellite
\cite{glast} should also give important data in the range where the
various final states cooperate to give a signal. The fate of the
produced electrons and positrons should also be investigated,
something that will need educated guesses for the magnetic field and
exact DM density distribution near the GC. In fact, a slightly
boosted NFW profile seems to predict a radio signal which is quite
close to that observed \cite{ber,aloisio}. We find further that the
direct annihilation into neutrinos will, in the KK model and with the
same halo parameters explaining the gamma-ray signal, give an event
rate corresponding to a few events per km$^2$ per year \cite{ber},
something that could eventually be detected in a neutrino telescope
with good view of the GC.

\section{Conclusions}

In this Letter we have presented the somewhat surprising result that
gamma rays radiated from charged leptons in $\B\B$ annihilations
(e.g., near the GC, but also in other overdense regions of the halo,
and maybe even in the diffuse extragalactic gamma-ray spectrum
\cite{BEU_01}) could be the most promising way to detect the effects
of KK DM and to possibly differentiate it from other candidates. We
have seen that the signature is a surprisingly hard spectrum, at high
energies several orders of magnitude larger than the previously
studied case of quark fragmentation alone, which extends all the way
to the mass of the DM particle and then drops rapidly. Even if the
present gamma-ray signal detected by H.E.S.S. will turn out to have
another cause, the sharp drop at $E_\gamma=m_{\B}$ is a nice signature
for that and other experiments to look for.

L.B. is grateful to the Swedish Science Research Council (VR) for
support. We acknowledge useful discussions with J.~Edsj\"o and
L.~Pieri.


\end{document}